\documentclass[cameraready]{Interspeech}
\usepackage[table]{xcolor}
\usepackage{tabularx}
\usepackage{amssymb}
\usepackage{xurl} 

\title{LISE : Listenable Interpretable Speaker Embeddings}

\author[affiliation={1}]{Xiaoliang}{Wu}
\author[affiliation={2}]{Chongxin}{Gan}
\author[affiliation={3}]{Ke}{Liu}
\author[affiliation={3}]{Peter}{Bell}
\author[affiliation={1}]{Jennifer}{Williams}

\address{
    $^1$ University of Southampton, United Kingdom \\
    $^2$ The Hong Kong Polytechnic University, Hong Kong SAR, China \\
    $^3$ University of Edinburgh, United Kingdom
}

\email{xiaoliang.wu@soton.ac.uk}


\usepackage{comment}

\begin{document}

\maketitle

\begin{abstract}
Deep neural network-based automatic speaker verification (ASV) systems achieve impressive performance but their embedding representations remain opaque, lacking a structured and perceptually verifiable explanation of the vocal characteristics they encode. Existing approaches either require annotation of speaker attributes or introduce alternative representations whose interpretability is unvalidated with listeners.
We propose \textbf{L}istenable \textbf{I}nterpretable \textbf{S}peaker \textbf{E}mbeddings (LISE), a label-free framework that decomposes pretrained speaker embeddings into a small set of components. This decomposition yields a structured representation that supports the analysis of what information has been encoded by speaker embeddings.
LISE preserves ASV performance with negligible EER degradation on x-vector and ECAPA-TDNN. Crucially, the interpretability of these components for human listeners is demonstrated through listening experiments, where participants distinguished speakers with 83.9\% accuracy.


\end{abstract}

\section{Introduction}
\label{sec:intro}

Modern automatic speaker verification (ASV) systems represent speaker identity using high-dimensional embeddings learned by deep neural networks~\cite{xvector,desplanques2020ecapa,conformer}. While highly effective for speaker discrimination, information encoded in these embeddings remains opaque, failing to provide a structured, perceptually verifiable account of what voice characteristics are captured. Consequently, interpretable decompositions could help reveal which vocal characteristics contribute to ASV~\cite{wu2024explainableSV}, enable the diagnosis of model biases~\cite{svbiases1,svbiases2}, and support applications such as controllable voice synthesis~\cite{lux2023controllable}. 

Prior work has explored two primary strategies for interpreting speaker embeddings, each involving trade-offs. One line of research associates speaker embeddings with predefined group-level attributes (e.g., age, accent) using probing classifiers \cite{wu2024explainableSV,chau2020,vove,huckvale2025interpretingdimensionsspeakerembedding} or disentanglement techniques \cite{williams2019disentangling,luu2022investigating}. Despite their effectiveness, these methods have three limitations. First, they require labeled attributes that are costly to annotate and raise privacy concerns. Second, categorical attributes cover only a subset of speaker variation -- they are inherently incapable of capturing other fine-grained vocal traits like breathiness or roughness that also contribute to speaker discrimination. 
Third, optimizing for attribute prediction may conflict with the objective of speaker verification, often resulting in severe performance degradation. For instance, Wu et al. \cite{wu2024explainableSV} found that using attribute classifiers on x-vector embeddings led to a substantial performance drop, with the equal error rate (EER) rising from 2.3\% to 17.0\%.


To eliminate label dependence, another line of work \cite{iben2022ba,iben2024extraction,iben2024forensic} reformulates speaker encoders for intrinsic interpretability by expanding embedding dimensionality and imposing binary constraints to obtain sparse representations. While verification performance is maintained (e.g., \cite{iben2024extraction} achieved comparable EER on x-vector), the interpretability claims remain unclear. Analysis typically focuses on correlations between individual dimensions and low-level acoustic cues such as formant structure and spectral shape~\cite{opensmile}, but correlation with acoustic cues does not guarantee that humans perceive the difference or find the dimension interpretable. This highlights a critical gap that existing methods do not simultaneously achieve label-free learning, performance preservation, and human-validated interpretability.


Sparse autoencoders (SAEs), have been used to explore the interpretability of text language models in the field of natural language processing (NLP) \cite{Subramanian2018SPINE,Cunningham2023SparseFeatures,Shi2025RouteSAE}, and offer a promising direction. By expanding embeddings to high dimensions while enforcing sparse activation, SAEs achieve two goals: (1) faithful reconstruction preserves task performance and (2) sparsity encourages each dimension to specialize. For instance, a dimension might activate strongly for words like "king, queen, prince," revealing an underlying concept of royalty. 

However, speaker embeddings differ structurally from word or sentence embeddings in ways that make SAEs unsuitable. Language contains thousands of discrete semantic units (words, entities, syntactic patterns), which naturally support high-dimensional sparse binary representations and correspond to concepts that are readily interpretable by humans. In contrast, the range of voice characteristics that humans can meaningfully describe is far more limited~\cite{ismail2024objective,vove}. 
Prior studies have identified only a few dozen commonly used voice descriptors—for instance, 24 descriptors in~\cite{ismail2024objective} and 44 in~\cite{vove}.
Those descriptors are also always subjective and vague. For example, in~\cite{vove}, voice descriptors include ``dark'', ``bright'', ``lively'' and ``sweet''. However, speaker variation can be expressed in a continuous rather than a binary form. A voice can range from ``dark'' to ``bright'', without a sharp boundary -- most fall somewhere in between. Moreover, feature meaning in NLP can be revealed by examining activating text, interpreting dimensions in speaker embeddings requires perceptual assessment. \textbf{We argue that dimension meaning is established when it corresponds to a recognizable auditory characteristic for human listeners.}



We propose LISE (Listenable Interpretable Speaker Embeddings) to address these challenges. LISE decomposes pretrained speaker embeddings into a limited set of components through unsupervised reconstruction. The approach follows three design principles tailored for speaker embeddings: (1) \textit{low dimensionality} to capture continuous structure rather than thousands of sparse features. (2) \textit{non-negative continuous weights} to represent graded perceptual differences without forcing binary decisions. (3) \textit{orthogonality} to ensure interpretable component independence. To further verify whether listeners can reliably distinguish speakers based on component weights, we conduct extensive listening experiments. In summary, we have made two contributions in this work:

\begin{enumerate}
\item \textbf{Label-free performance-preserving structured decomposition of speaker embedding:} 
We propose LISE, a post-hoc decomposition of pretrained speaker embeddings into a compact set of orthogonal, non-negative components without attribute supervision. This decomposition maximally preserves speaker verification performance on Vox1-O.

\item \textbf{Human perceptual validation of component interpretability:} 
We design a listening-based evaluation protocol to assess the perceptual validity of the learned components. Results indicate that LISE components are reliably discriminated by human listeners.
\end{enumerate}

\section{Listenable interpretable speaker embeddings (LISE)}
SAEs work well for word and language embeddings by identifying discrete semantic features~\cite{Subramanian2018SPINE,Cunningham2023SparseFeatures,Shi2025RouteSAE}. However, as discussed in the Section~\ref{sec:intro}, speaker embeddings have fundamentally different structure: voice characteristics that listeners can describe are limited, and speaker variation is continuous rather than categorical. This implies that directly applying sparse high-dimensional decomposition is impractical.

\begin{figure}[ht]
    \centering
    \includegraphics[width=\linewidth]{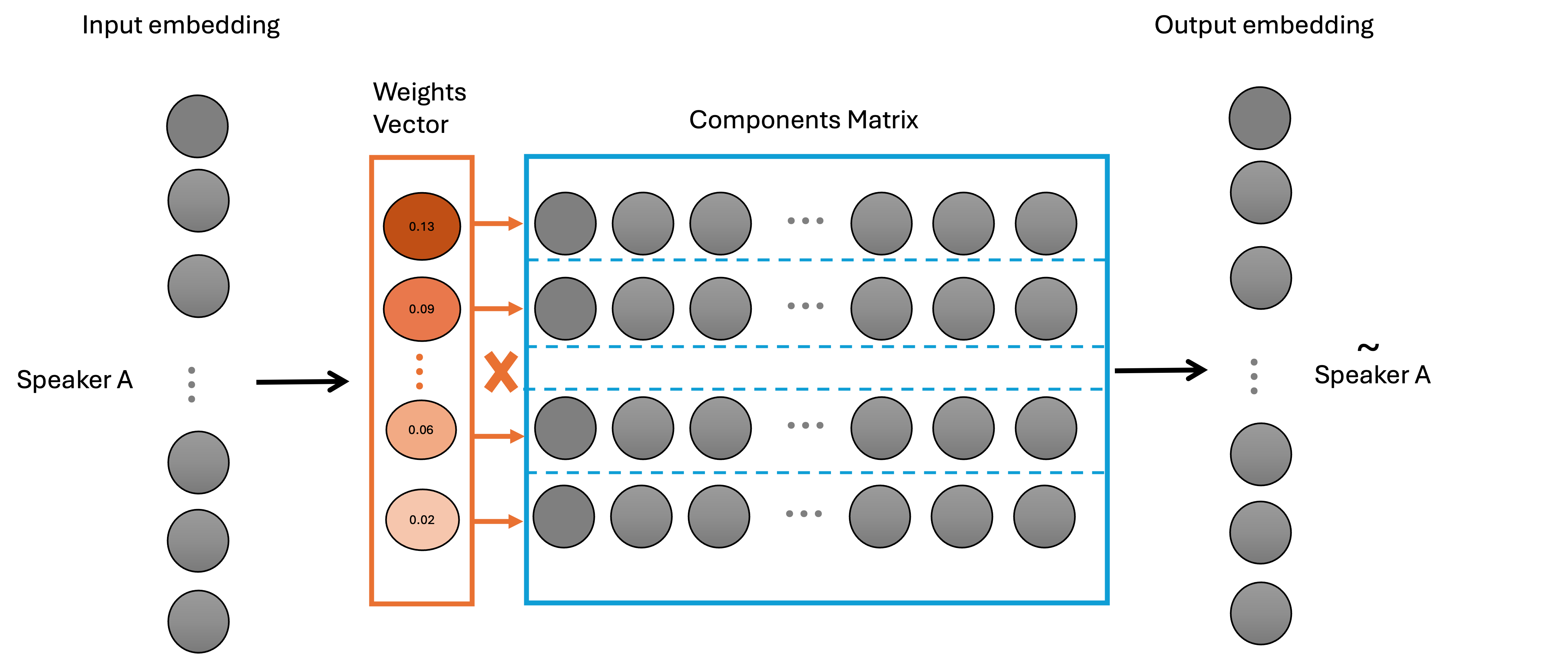}
    \caption{A speaker embedding is decomposed into $K$ orthogonal components with non-negative weights.}
    \label{fig:lise_framework}
\end{figure}

As shown in Figure~\ref{fig:lise_framework}, LISE decomposes the speaker embedding $\bm{e} \in \mathbb{R}^d$, which is produced by a pretrained speaker encoder, into $K$ interpretable components. Given a component matrix $\bm{W} \in \mathbb{R}^{d \times K}$ and weights $\bm{c} \in \mathbb{R}^K$, the embedding is reconstructed as
\begin{equation}
\hat{\bm{e}} = \bm{W}\bm{c}.
\end{equation}

\textbf{Low-dimensional structure.} Unlike SAEs in NLP adopting thousands of sparse features, LISE imposes a relatively strong constraint: $K \ll d$, typically using $K \in 5$--$50$ components. This avoids dispersing continuous voice characteristics across many weakly-activated dimensions. Unlike SAEs, low dimensionality is intentional. It respects the natural structure of speaker variation rather than working against it. The impact of $K$ is analyzed empirically in Section~\ref{sec:verification}.

\textbf{Non-negative additive contributions.} To ensure interpretability, LISE enforces non-negativity: $\mathbf{c} \ge \mathbf{0}$. If both positive and negative weights were allowed, reconstruction could involve subtractive interactions where one component partially cancels another, making interpretation ambiguous. Non-negative weights ensure each component's contribution directly reflects the individual components' contributions. Moreover, continuous weights enable gradual interpretation: small changes in voice characteristics produce small weight adjustments rather than forcing binary present/absent decisions.

\textbf{Orthogonality for component independence.} LISE learns $\bm{W}$ and $\bm{c}$ by optimizing
\begin{equation}
\mathcal{L} = \lVert \bm{e} - \hat{\bm{e}} \rVert_2^2 + \lambda\, \lVert \bm{W}^\top \bm{W} - \mathbf{I} \rVert_F^2,
\label{eq:LISE}
\end{equation}

where the first term ensures reconstruction fidelity and the second promotes orthogonality. Reconstruction alone permits entangled components that jointly minimize error but are difficult to interpret individually. In high-dimensional sparse models, sparsity ensures each sample activates only a small subset of features, naturally reducing redundancy. In LISE's low-dimensional setting, orthogonality serves an analogous role: it reduces redundancy and ensures components capture distinct axes of variation that can be examined independently.

\section{Experimental setup}
We describe our datasets, training setup, baseline methods, and listening study protocols for validating LISE in terms of ASV performance and human perceptual studies.

\subsection{Datasets and embedding extractors}
We use VoxCeleb2~\cite{nagrani2017voxceleb} training set (5,994 speakers, approximately 1.1M utterances) to train LISE. Speaker embeddings are extracted using two pretrained models from SpeechBrain~\cite{ravanelli2021speechbrain}: x-vector (512-dim) and ECAPA-TDNN (192-dim). For each speaker, we average utterance-level embeddings to obtain speaker-level representations. The embedding extractors remain frozen. LISE is applied post-hoc without retraining. Speaker verification performance is evaluated on VoxCeleb1-O~\cite{nagrani2017voxceleb} using Equal Error Rate (EER) derived from the cosine similarity score on reconstructed embeddings. 

\subsection{LISE training}
We optimize Equation~\ref{eq:LISE} using non-negative least squares~\cite{LawsonHanson1995Solving} with orthogonality regularization ($\lambda = 0.05$). Training runs for 200 epochs on 5,994 speaker embeddings using a single NVIDIA 2080Ti GPU. We train LISE for different numbers of components $K \in \{5, 10, 15, 20, 25, 30, 35, 40, 45, 50\}$ and select $K=35$ based on the verification performance trade-off (detailed in Section~\ref{sec:verification}).

\subsection{Baseline methods}
We compare LISE against three methods: (1) \textbf{Principal Component Analysis} (PCA) \cite{Hotelling1933PCA} --- a classic unsupervised linear decomposition that, unlike LISE, allows negative weights; (2) \textbf{Luu et al.}~\cite{luu2022investigating}: adversarial training with attribute (gender,nationality) supervision; (3) \textbf{Iben et al.}~\cite{iben2024extraction}: high-dimensional binary embeddings with 0/1 activations. 
All methods are trained on the same VoxCeleb2 and evaluated on VoxCeleb1-O. Original embeddings without involving decomposition serve as the performance upper bound.



\subsection{Perceptual validation protocol}
\label{sec:perceptual_protocol}
To test whether LISE components are interpretable to human listeners, we design a discrimination task based on the following logic: if a component captures a coherent perceptual property, speakers with high weights on that component should exhibit similar perceptual voice characteristics, and listeners should reliably distinguish them from speakers with low weights.

\begin{figure}[ht]
    \centering
    \includegraphics[width=0.55\linewidth]{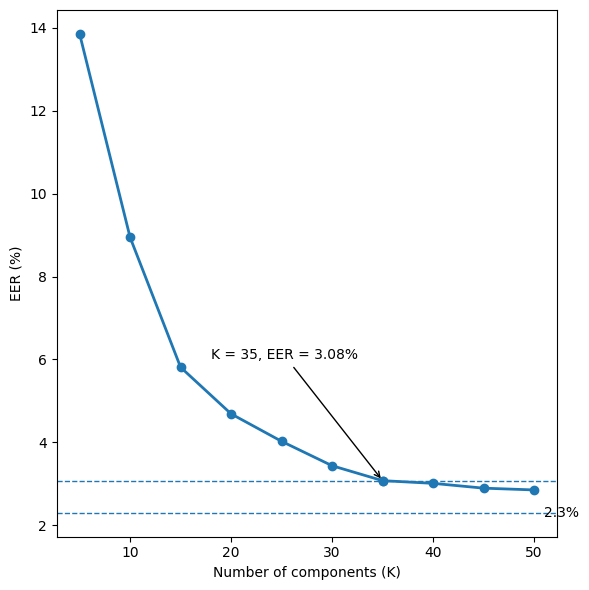}
    \caption{Effect of the number of LISE components $K$ on speaker verification performance (x-vector).}
        \vspace{-15pt}
    \label{fig:k_sensitivity}
\end{figure}

For each component $k$, we rank 5,994 speakers by their component weight and form three sets. \textit{Type A} consists of the top 3 speakers with highest weights. \textit{Non-Type A} consists of the bottom 3 speakers with lowest weights. \textit{Candidates} are speakers ranked 4th--6th (positive) and 4th--6th from bottom (negative). Positive and negative candidates are balanced in equal proportions. The listening task proceeds in two stages. First, in familiarization stage, participants listen to one file from each reference speaker (6 files total: 3 \textit{Type A} and 3 \textit{Non-Type A}). They are explicitly informed which group each speaker belongs to, allowing them to learn the perceptual contrast. Second, in testing, participants judge 2--3 candidate files, deciding whether each belongs to \textit{Type A} or \textit{Non-Type A}. Across all 35 components, this yields $95$ judgments per participant.


We analyze accuracy at three complementary levels. Accuracy is measured as the percentage of correct judgments:
\begin{equation}
\text{Accuracy} = \frac{\text{correct judgments}}{\text{total judgments}}
\label{eq:accuracy}
\end{equation}
Overall accuracy across all participants and judgments validates whether LISE's decomposition is generally interpretable. Per-component accuracy identifies which components are more reliably interpretable to listeners. Per-participant accuracy assesses whether interpretability is consistent across different listeners.

We compare three 35-dimensional unsupervised methods that preserve verification performance well: LISE, PCA, and Iben et al.~\cite{iben2024extraction}. Notably, for 512-high-dimensional representation from Iben et al., we select the 35 most discriminative dimensions for fair comparison in the listening task. We use an iterative ablation procedure: we remove each dimension one at a time and measure the resulting EER on VoxCeleb1-O. We retain the 35 dimensions where removal causes the largest increase in EER, as these dimensions are most important for speaker discrimination. All three methods then use identical reference set formation, familiarization protocol, balanced candidate selection, and testing procedure. For each method, the same 25 participants (all above 18 years old, self-reported normal hearing) completed the full listening study for all 35 components of their assigned method.



\section{Results discussion}

This section presents experimental evaluation of LISE from two aspects. First, verification performance -- does LISE preserve discriminative capability? We report performance on a speaker verification task. Second, perceptual interpretability -- can humans reliably distinguish components?. We validate interpretability using a listening test.  

\subsection{Speaker verification performance}
\label{sec:verification}

We first examine how the number of components $K$ affects verification performance. Figure~\ref{fig:k_sensitivity} shows EER results on VoxCeleb1-O when $K$ varies. It is observed that EER decreases rapidly until $K \approx 35$, after which it plateaus. When $K=35$, LISE achieves 3.08\% EER for x-vector (original: 2.30\%) and 2.10\% for ECAPA-TDNN (original: 1.80\%). This demonstrates that most speaker-discriminative information is captured by a small set of additive components. We select $K=35$ as the smallest dimensionality where performance stabilizes, balancing interpretability and verification performance.

\begin{table}[t]
\centering
\caption{Speaker verification performance (EER\%) on VoxCeleb1-O across different interpretability methods. Lower is better. Original performance is the upper limit. LISE results on full, 75\%, and 50\% of the training dataset.}
\label{tab:performance}
\footnotesize
\begin{tabular}{lcc}
\toprule
\textbf{Method} & \textbf{x-vector}$\downarrow$ & \textbf{ECAPA-TDNN}$\downarrow$ \\
\midrule
Original & 2.30 & 1.80 \\
\midrule
Luu et al.~\cite{luu2022investigating}  & 6.70 & -- \\
Iben et al.~\cite{iben2024extraction}  & 3.34 & 2.50 \\
PCA  & \textbf{3.02} & 3.28 \\
\midrule
\rowcolor{green!15}
LISE (ours, full data) & 3.08 & \textbf{2.10} \\
\rowcolor{green!15}
LISE (ours, 75\% dataset) & 3.13 & 2.15 \\
\rowcolor{green!15}
LISE (ours, 50\% dataset) & 3.23 & 2.18 \\
\bottomrule
\end{tabular}
\end{table}

Table~\ref{tab:performance} compares LISE against other methods. LISE effectively preserves speaker verification performance, remaining competitive when compared with other unsupervised methods (PCA and Iben et al.~\cite{iben2024extraction}\footnote{The original paper~\cite{iben2024extraction} used a stronger baseline. Here we re-implemented the method using SpeechBrain embeddings for consistency.}). On x-vector, LISE (3.08\%) performs comparably to PCA (3.02\%). On ECAPA-TDNN, LISE (2.10\%) also achieves near-original performance (1.80\%) and outperforms both PCA (3.28\%) and Iben et al. (2.50\%).

In contrast, prior supervised attribute-based methods~\cite{luu2022investigating} from Luu et al. suffer severe performance degradation (6.70\% on x-vector). This occurs because optimizing for predefined categorical labels (gender or nationality) discards the fine-grained discriminative information that is critical to distinguish individual speakers -- many speakers could share identical or similar labels yet sound completely different. LISE avoids this pitfall by learning interpretable structure directly from the embedding space without requiring attribute labels or imposing predefined categories.

We also evaluate LISE's robustness on data scarcity. Green zone in the Table~\ref{tab:performance} shows LISE performance when trained on only 75\% or 50\% of utterances within the dataset. The degradation is minimal: at 75\% dataset, LISE achieves 3.13\% and 2.15\% on x-vector and ECAPA respectively; at 50\%, it achieves 3.23\% and 2.18\%. This robustness reflects a key property of the pretrained embeddings: x-vector and ECAPA-TDNN are trained on the full VoxCeleb2 dataset, so they already encode sufficient speaker-level information to support stable decomposition even with reduced utterance diversity.
\vspace{-5pt}

\begin{figure}[ht]
    \centering
    \includegraphics[width=\columnwidth]{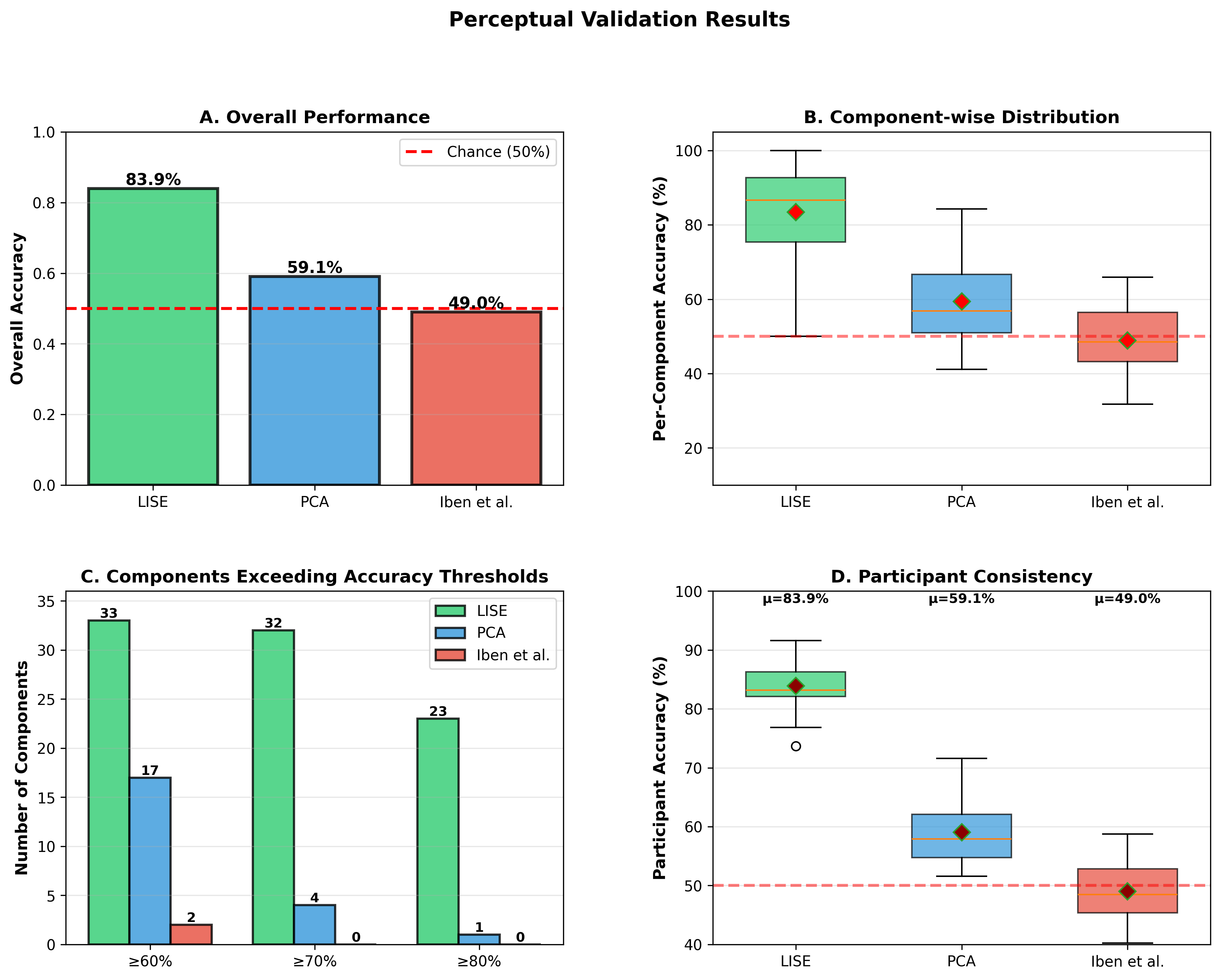}
    \caption{Perceptual validation comparing LISE, PCA, and Iben et al.~\cite{iben2024extraction} across four aspects: (A) overall accuracy, (B) per-component accuracy, (C) number of components exceeding accuracy thresholds, (D) participant consistency.}
    \label{fig:perceptual_validation}
    \vspace{-15pt}
\end{figure}

\subsection{Perceptual validation}
Preserving the SV task performance alone (from Section~\ref{sec:verification}) does not guarantee that decomposed components are interpretable to humans. To assess whether listeners can reliably distinguish speakers based on component weights, we conducted listening experiments (see Section~\ref{sec:perceptual_protocol} for task design details).

\textbf{Overall accuracy.} LISE achieves 83.9\% accuracy across all 
25 participants (Figure~\ref{fig:perceptual_validation}A). In comparison, PCA achieves 59.1\% and Iben et al.\ achieves 49.0\%. LISE’s substantial performance gains -- surpassing PCA by 24.8 percentage points and Iben et al. by 34.9 percentage points -- demonstrate that its components are considerably more discriminable to listeners.

These results align with our design expectations. PCA is mathematically interpretable through orthogonal decomposition but lacks non-negativity constraints, allowing components to offset each other and making it harder for listeners to identify which component causes speaker differences. LISE's non-negative weights ensure each component's contribution is independent and clear. Prior work~\cite{iben2024extraction} preserved verification performance but also spread speaker variation across 512 sparse binary dimensions. In that case, a dimension would represent something so specific as "F0 is 30 Hz or not," and that distinction is difficult to perceive. LISE's low dimensionality with continuous weights allows each component to capture more substantial variation, making differences perceptually salient.

\textbf{Per-component accuracy.} Examining individual component accuracy reveals different patterns across methods. As presented in Figure~\ref{fig:perceptual_validation}B, LISE achieves the best results: mean 83.4\%, median 84.8\%, with most components between 73\% -- 90\%. This indicates interpretability is broadly distributed, not concentrated in a few good components. In contrast, PCA exhibited lower performance: mean 59.4\%, median 56.3\%, range from 51\% to 69\%. The method by Iben et al. performed most poorly: mean 48.9\%, median 46.7\%, with many components at or below chance. This clearly suggests that most components in LISE are more interpretable than the other two methods.

\textbf{Interpretable components count.} Figure~\ref{fig:perceptual_validation}C then shows this pattern by counting components above accuracy thresholds. At 70\% accuracy, LISE has 32 out of 35 (94.2\%), compared to 4 for PCA (11.4\%) and 0 for Iben et al. At 80\% accuracy, LISE maintains 23, while PCA drops to 1. Across both thresholds, LISE provides substantially more reliably usable components.

\textbf{Participant consistency.} Individual participants may differ in hearing ability, familiarity with English, and attention. To check whether results depend on particular listeners, we examine each participant's accuracy (Figure~\ref{fig:perceptual_validation}D).
LISE shows consistent high performance wherein all participants achieve 73.7--91.6\% accuracy. PCA shows lower and more variable results with 51.6--71.6\% accuracy. Prior work from Iben et al.\ clusters near chance: 40.2\%--58.8\%.

\textbf{Qualitative observations.} During the listening tests, some participants spontaneously provided natural language descriptions of the perceived commonalities among speakers within each component's \textit{Type A} group. 
Table~\ref{tab:top_components} shows the five highest-ranked components based on listener discrimination accuracy, illustrating the voice characteristics captured by LISE. Complete audio samples for all 35 components are available online\footnote{\url{https://sites.google.com/view/components-samples/home}}. For each component, we provide speech audio from the three speakers with the highest component weights (used as \textit{Type A} references in the listening test), together with participant descriptions. Readers are encouraged to listen to these examples and form their own perceptual interpretations.

\begin{table}[t]
\centering
\caption{Representative participant descriptions of highly interpretable LISE components, ordered by discrimination accuracy.}
\label{tab:top_components}
\footnotesize
\begin{tabular}{lcc}
\toprule
\textbf{Component} & \textbf{Participant Description} & \textbf{Accuracy} \\
\midrule
C9  & Slower-paced female & 100\% \\
C27 & Deep resonant voice & 97\% \\
C21 & Young female, rapid tempo & 97\% \\
C18 & Bright youthful female & 94\% \\
C19 & Male, creaky voice & 94\% \\
\bottomrule
\end{tabular}
\end{table}


\textbf{Future work: listenable component prototypes.} Because LISE components reside in the original speaker embedding space, they can directly condition speaker-aware TTS systems. As a preliminary exploration, we used SpeechT5~\cite{speecht5} to synthesize speech from component vectors, producing audible ``prototypes'' for each component. 
These synthesized audio examples are available online\footnote{\url{https://sites.google.com/view/lise-prototypes/home}}. Systematic evaluation of TTS-based validation remains future work, but points toward promising applications in controllable voice synthesis where users manipulate interpretable vocal characteristics, rather than opaque embedding coordinates. Although VoxCeleb is an assumed English-only dataset, some speakers exhibit non-English languages. Consequently, a small number of components (3/35) capture language patterns rather than speaker-intrinsic vocal characteristics. Future work should validate LISE on strictly monolingual datasets, and also explore potential use-cases for interpretable embeddings built systematically upon multilingual data, such as speech-speech translation or accent control.
\vspace{-5pt}



\section{Conclusion}

We introduced LISE, a label-free framework to decompose pretrained speaker embeddings into interpretable components while preserving verification performance (3.08\% EER for x-vector, 2.10\% for ECAPA-TDNN) 
A key contribution of this paper is our perceptual validation. Listening experiments show that LISE components are genuinely interpretable to humans. Our evaluation shows that LISE achieves superior perceptual interpretability compared to unsupervised methods demonstrated by higher overall accuracy
, more concentrated per-component performance, and consistent participant perception.  Future work will explore applications in controllable voice synthesis, bias diagnosis, and auditory speaker analysis.

\section{Acknowledgements}
This work was supported by the Engineering and Physical Sciences Research Council (EPSRC) through the National Edge AI Hub for Real Data: Edge Intelligence for Cyberdisturbances and Data Quality (EP/Y028813/1) and Responsible AI UK (EP/Y009800/1).

\section{Generative AI tools disclosure}
Generative artificial intelligence tools were used solely to assist with language editing and clarity of presentation. All research ideas, methodology, experiments, and interpretations were conceived and carried out by the authors, who take full responsibility for the originality, validity, and integrity of the work.

\bibliographystyle{IEEEtran}
\bibliography{mybib}

\end{document}